\documentclass[fleqn]{aa61}
\usepackage{graphicx,natbib,txfonts,url,amssymb}
\bibpunct{(}{)}{;}{a}{}{,}

\newcommand{\figureone}[3]{%
\begin{figure}[tbp]
\begin{center}
\includegraphics[width=88mm]{#1}
\caption{#3}
\label{#2}
\end{center}
\end{figure}
}

\newcommand{\figuretwo}[3]{%
\begin{figure*}[tbp]
\begin{center}
\includegraphics[width=180mm]{#1}
\caption{#3}
\label{#2}
\end{center}
\end{figure*}
}

\newcommand\CIV{\mbox{C\,\sc{iv}}} 
\newcommand\Ha{\mbox{H$\alpha$}}
\newcommand\CaIIHK{\mbox{Ca\,\sc{ii}\,H\&K}}

\newcommand\FeXIX{\mbox{Fe\,\sc{ix}/\sc{x}}} 
\newcommand\mA{m\AA}

\begin{document}

%%% HEADER %%%%%%%%%%%%%%%%%%%%%%%%%%%%%%%%%%%%%%%%%%%%%%%%%%%%%%%%%%%%%%%%
\title{Dynamic fibrils in \Ha\ and \CIV}
\titlerunning{}
\author{A.G.~de~Wijn\inst{1}
		\and
		B.~De~Pontieu\inst{2}}
\institute{Sterrekundig Instituut, 
           Utrecht University, 
           Postbus~80\,000, 3508~TA~Utrecht, The~Netherlands\\
           \email{A.G.deWijn@astro.uu.nl}
           \and
		   Lockheed Martin Solar and Astrophysics Laboratory,
		   3251 Hanover Street, Org. ADBS, Building 252, Palo Alto,
		   California 94304, USA\\
		   \email{bdp@lmsal.com}}
\date{Received 20 June 2006 / Accepted 5 September 2006}

\abstract%
{}%
{To study the interaction of the solar chromosphere with the transition region, in particular active-region jets in the transition region and their relation to chromospheric fibrils.}%
{We carefully align image sequences taken simultaneously in \CIV\ with the Transition Region and Coronal Explorer and in \Ha\ with the Swedish 1-meter Solar Telescope.
We examine the temporal evolution of ``dynamic fibrils'', i.e., individual short-lived active-region chromospheric jet-like features in \Ha.}%
{All dynamic fibrils appear as absorption features in \Ha\ that progress from the blue to the red wing through the line, and often show recurrent behavior.
Some of them, but not all, appear also as bright features in \CIV\ which develop at or just beyond the apex of the \Ha\ darkening.
They tend to best resemble the \Ha\ fibril at $+700$~\mA\ half a minute earlier.}%
{Dynamic chromospheric fibrils observed in \Ha\ regularly correspond to transition-region jets observed in the ultraviolet.
This correspondence suggests that some plasma associated with dynamic fibrils is heated to transition-region temperatures.}

\maketitle

%%%%%%%%%%%%%%%%%%%%%%%%%%%%%%%%%%%%%%%%%%%%%%%%%%%%%%%%%%%%%%%%%%%%%%%%%%%
\section{Introduction}\label{sec:introduction}
%%%%%%%%%%%%%%%%%%%%%%%%%%%%%%%%%%%%%%%%%%%%%%%%%%%%%%%%%%%%%%%%%%%%%%%%%%%
In this paper, we study active-region fibrils in the solar chromosphere and transition region.
The context is the dynamical structure of the transition region as the interface between the chromosphere and the corona.
It is highly complex due to the fine structure of the chromosphere, which is dominated on small scales by jet-like features observed in \Ha\ as spicules at the limb, and as active-region fibrils and quiet-sun mottles on the disk.
They bring cold ($\sim$$10^4$~K), dense plasma up to greater heights than expected from hydrostatic equilibrium
	\citep[e.g.,][]{1995ApJ...450..411S}. % Suematsu, disk spicules in network
The understanding of the formation and behavior of these structures is limited, although they appear to be key agents in the momentum and energy balances of the outer solar atmosphere
	\citep{2004A&A...424..279T}. % Tsiropoula energy & mass balance
Even the relationships between these different features remain in debate
	(see the still valid observational reviews by \cite{1968SoPh....3..367B,1972ARA&A..10...73B}, % Beckers, spicule reviews
	the theoretical review by \cite{2000SoPh..196...79S}, % Sterling, spicule models review
	and the more recent analyses by \cite{2003A&A...402..361T,2004A&A...423.1133T} and references therein). % Tziotziou, mottles

Solar images taken in EUV lines also show jet-like structures in the chromosphere and transition region.
Some EUV jets extend much further out from the limb than \Ha\ spicules, and have much longer lifetimes.
However, it has been established that many EUV and \Ha\ features tend to have similar lengths and dynamics
	\citep[e.g.,][]{1983ApJ...267..825W, % Withbroe EUV limb emission =? spicules
	1983ApJ...267L..65D, % Dere et al EUV spicules
	1986ApJ...305..947D, % Dere et al on disk EUV jets
	1987SoPh..114..223D, % Dere et al EUV spicules subresolution structure
	1998A&A...334L..77B, % Budnik spicule evaporation from SUMER
	2000A&A...360..351W, % Wilhelm spicules from SUMER
	2005A&A...438.1115X}. % Xia time series spicules by SUMER
The similarities between a class of EUV and \Ha\ features suggests that they are different manifestations of the same phenomenon, possibly with EUV jets being hot sheaths around cool \Ha\ interiors.
However, a direct one-to-one correspondence has not yet been established.

In the present analysis, we concentrate on short-lived active-region fibrils observed in \Ha.
These have lifetimes of only a few minutes.
We call them ``dynamic fibrils''.
Longer and more static fibrils are not the subject of this study.
	\cite{2003ApJ...590..502D} % De Pontieu, moss vs C IV vs Halpha
reported spatial coincidence of emission features in \CIV\ lines and dynamic \Ha\ fibrils observed in the red wing at $\Delta\lambda=700$~\mA\ from line center.
The dynamic fibrils in \Ha\ correspond well with dark inclusions observed in the \FeXIX\ EUV passband around $171$~\AA\ of the Transition Region and Coronal Explorer
	\citep[TRACE,][]{1999SoPh..187..229H} %T Handy etal, TRACE mission
in so-called mossy plage, where ``moss'' describes the finely textured and dynamical appearance of such plage in $171$-\AA\ image sequences
	\citep{1999SoPh..190..409B, % Berger moss
	1999SoPh..190..419D, % De Pontieu moss at high t res
	2004Natur.430..536D}. % BDP fibrils in Nature
Here, we expand on
	\cite{2003ApJ...590..502D} % De Pontieu, moss vs C IV vs Halpha
by examining the temporal evolution of the dynamic \Ha fibrils and their \CIV\ counterparts.
The data are described in Sect.~\ref{sec:observations}, the analysis in Sect.~\ref{sec:analysis}, and the results are discussed in Sect.~\ref{sec:discussion} and summarized in Sect.~\ref{sec:conclusion}.

%%%%%%%%%%%%%%%%%%%%%%%%%%%%%%%%%%%%%%%%%%%%%%%%%%%%%%%%%%%%%%%%%%%%%%%%%%%
\section{Observations and data reduction}\label{sec:observations}
%%%%%%%%%%%%%%%%%%%%%%%%%%%%%%%%%%%%%%%%%%%%%%%%%%%%%%%%%%%%%%%%%%%%%%%%%%%

% rubber: depend bm_sample2003a.eps
\figureone{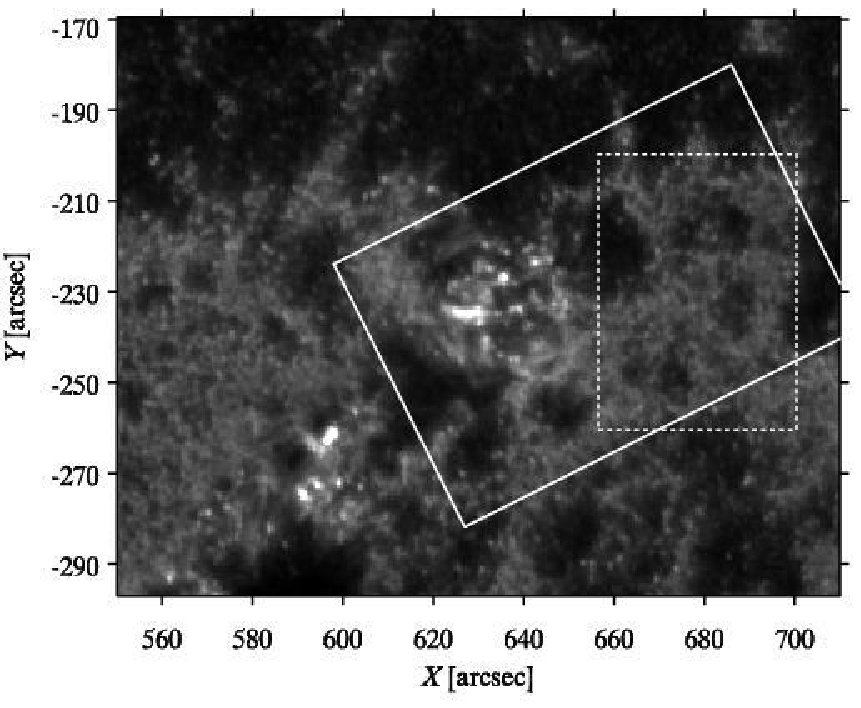}{fig:sample2003a}{%
A sample $1600$-\AA\ image cutout.
$X$ and $Y$ are the standard heliocentric coordinates with solar north up used by TRACE.
The SST field of view is indicated by the solid contour.
It contains active region AR10380 with a few small sunspots.
The dashed contour specifies the cutout used in Fig.~\ref{fig:sample2003b}.}

We analyze image sequences of NOAA active region 10380 recorded simultaneously in \Ha\ and in \CIV on June~16 2003 between 8:02 and 9:07~UT at a viewing angle of 48~degrees ($\mu=0.66$).
AR10830 was a sizable active region that decayed rapidly during its transition over the disk, appearing as a large complex of sunspots and disappearing as mostly plage.
It still contained several small sunspots at the time of observation.
The Swedish 1-m Solar Telescope
	\citep[SST,][]{2003SPIE.4853..341S} % Scharmer, SST
was used to take images at four wavelengths in \Ha, while co-pointed imaging was performed by TRACE in its three ultraviolet (UV) passbands at $1550$, $1600$, and $1700~\AA$.

At the SST, the SOUP tunable Lyot filter
	\citep[Solar Optical Universal Polarimeter,][]{1986AdSpR...6..253T} % Title, SOUP
was used for narrow-band imaging at $\Delta\lambda=-700$, $-350$, $+350$, and $+700$~\mA\ from the \Ha\ line center with $130$-\mA\ FWHM bandwidth.
These passbands sample the chromosphere, though there is a contribution from the photosphere at least in the $\pm700$-\mA\ passbands due to the double-peaked contribution function of the \Ha\ line wing
	\citep{2006A&A...449.1209L}. % Leenaarts Halpha wing
The images consist of $1534\times1032$ square pixels of $0.0655\arcsec$ size.
Bursts of images were recorded at each sampling wavelength during seven seconds before tuning the SOUP filter to the next wavelength.
The frame rate was $1.6$~fps at $\pm700$~\mA\ with $115$-ms exposure, and $1.2$~fps at $\pm350$~\mA\ with $200$-ms exposure.
Thanks to the use of the SST's adaptive-optics system, the data quality does not suffer too much from these long exposure times.
The three images of each burst with the highest contrast were stored to disk.

In the post-processing all stored images were first corrected for wavelength-dependent dark current and flat field.
In order to reduce seeing effects caused by the Earth's atmosphere, each set of three images was subsequently processed with the multi-frame blind deconvolution technique developed by
	\cite{2002SPIE.4792..146L} % Lofdahl MFBD
using the implementation of Van~Noort\footnote{\url{http://www.solarphysics.kva.se/mfbd}}
	\citep[cf.][]{2005SoPh..228..191V}. % Van Noort, MFBD
The resulting images reach better than $0.25\arcsec$ resolution.
The four \Ha\ sequences have a cadence of $55\pm5$~s, with an average time delay of $12\pm2$~s between successive wavelength positions, and $18\pm5$~s from the last red back to the first blue position.
They were rotated to the same orientation as the TRACE images with solar north up.
Image displacements were then measured in multiple steps described below.
These displacements were subsequently used to re-sample the de-rotated images onto an aligned grid in a single step in order to avoid repeated spatial interpolation.
We began by computing the displacements between successive images per wavelength.
Next, we corrected the $-700$-\mA\ sequence for residual linear drift by aligning the first and last images, and adjusting the computed displacements between these two.
The first and last images of the other passbands were then aligned to the $-700$-\mA\ ones using the wavelet technique of
	\cite{1999SoPh..190..419D}. % De Pontieu moss at high t res
Finally, we used the computed offsets to shift the de-rotated images using cubic-spline interpolation to create four accurately aligned sequences of 78~images each.

Visual inspection of the sequences shows rapid changes in the mean full-field intensity.
We suspect that the SOUP filter was not always precisely tuned to the wavelength specified.
Since we study the morphology of structures here, this amplitude variation and the small tuning error do not affect our results.

% rubber: depend bm_sample2003b.eps
\figuretwo{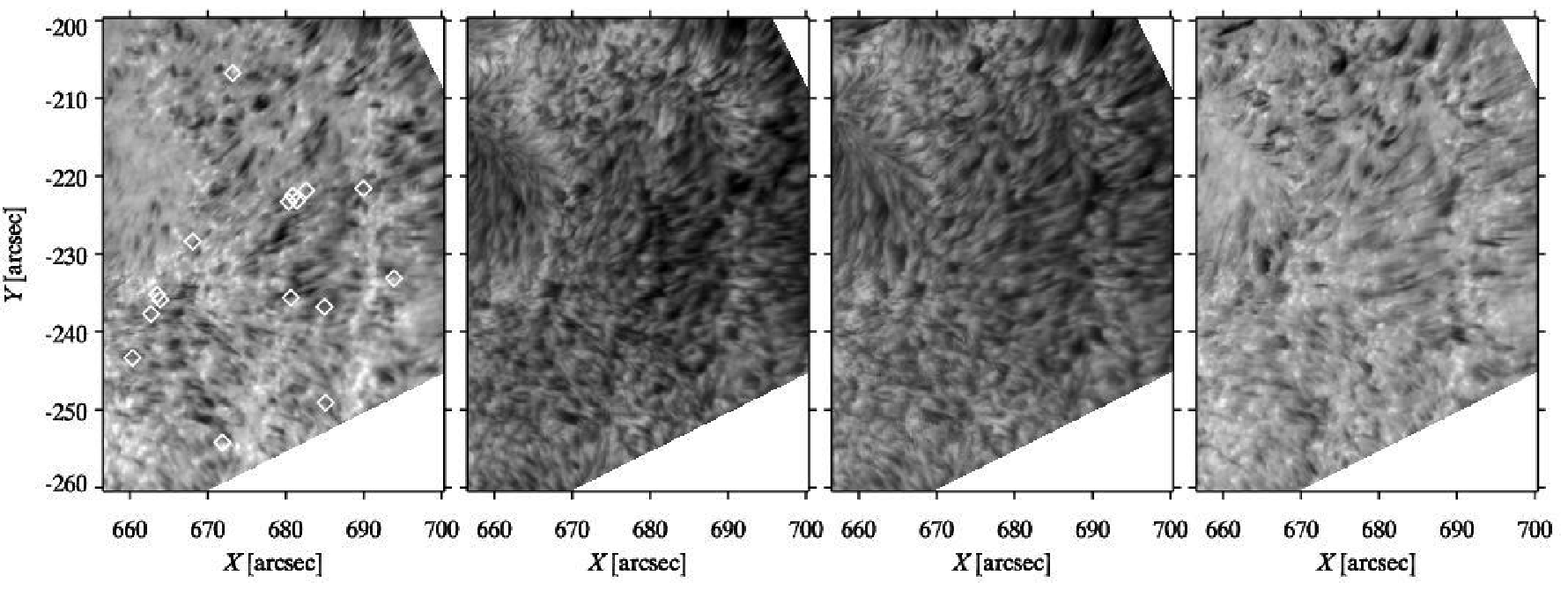}{fig:sample2003b}{%
Examples of \Ha\ image cutouts sampling the \Ha\ line from left to right at $-700$, $-350$, $+350$, and $+700$~\mA.
This part of the SST field of view shows many small dynamic fibrils that appear and disappear during the sequence.
The outer-wing images sample deeper layers than the inner-wing ones, showing dark chromospheric structures over a bright photospheric background.
The diamonds in the $-700$~\mA\ image mark the positions where fibrils occur that show associated \CIV\ emission at some time.}

% rubber: depend bm_spicsample.eps
\figuretwo{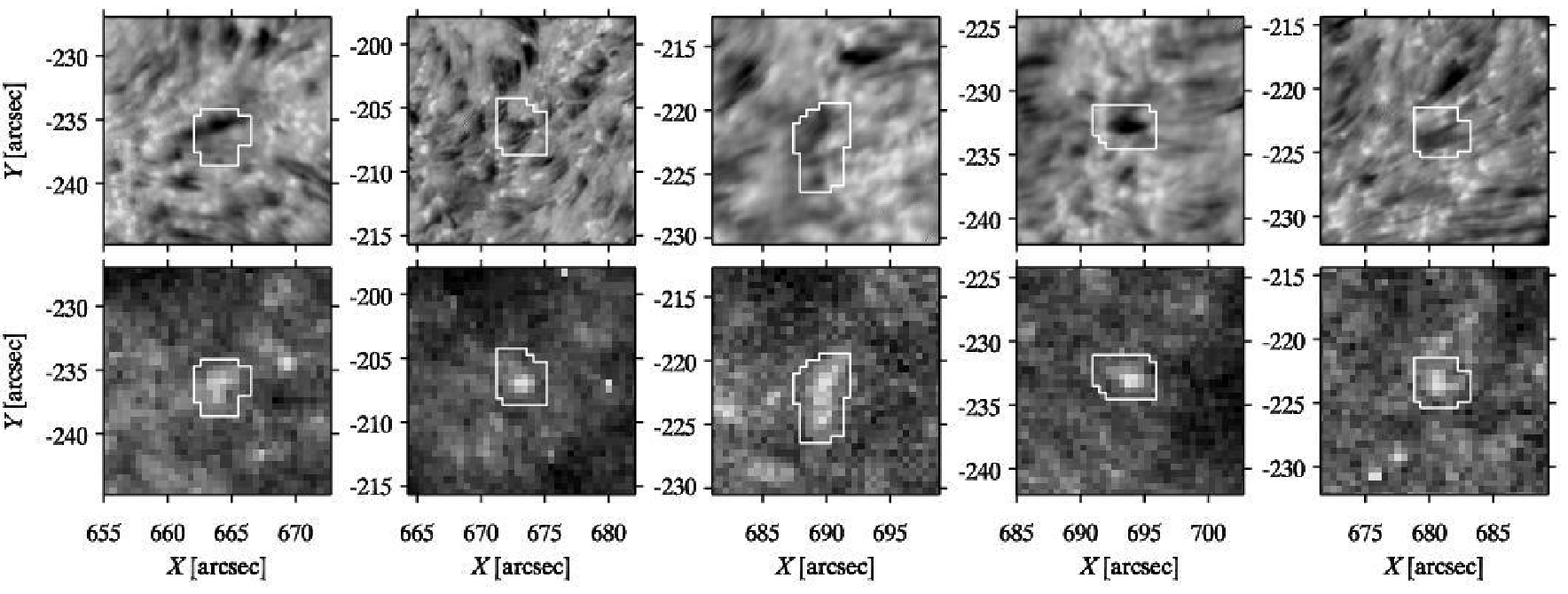}{fig:spicsample}{%
Five sample \Ha\ fibrils of 16 identified with cospatial-cotemporal \CIV\ emission.
\emph{Top row}: \Ha\ $+700$~\mA\ cutouts.
\emph{Bottom row}: \CIV\ cutouts.
In each case, excessive \CIV\ brightness coincides spatially with morphologically similar dark features in \Ha $+700$~\mA.
The white contours enclose the area over which the Pearson correlation coefficients discussed in Sect.~\ref{sec:pearson} are computed.}

The TRACE UV bandpasses labeled $1550$, $1600$, and $1700$~\AA\ together sample the low and upper chromosphere as well as the transition region.
In these observations TRACE was programmed to record images sequentially in these three wavelengths at a cadence of $20\pm1$~s.
The angular resolution is $0.5\arcsec$ per pixel, the image size $640\times512$ square pixels.
The images were corrected for dark current and were flat-fielded with the SolarSoft procedure \texttt{trace\_prep} described in the TRACE Analysis Guide\footnote{\url{http://moat.nascom.nasa.gov/~bentley/guides/tag}}.
The images were subsequently aligned with high precision using Fourier cross-correlation in three steps.
In the first step, the $1600$-\AA\ images were aligned in consecutive pairs, and each $1550$ and $1700$-\AA\ image was subsequently aligned with the $1600$-\AA\ image that is closest in time.
The second step involves alignment of each of the resulting images with respect to a running average over nine images.
The latter were selected from the output of the first step for $1550$, $1600$, and $1700$~\AA\ combined such as to be closest in time, and smoothed with a $5\times5$-pixel spatial averaging.
This procedure avoids copying alignment errors from one sequence to another, as would occur when aligning to a single ``master'' image sequence
	\citep{2005A&A...430.1119D}.
Finally, each original image was re-sampled to the aligned grid.
The single interpolation step ensures minimal degradation of the image quality.

The $1550$ and $1600$-\AA\ passbands include the \CIV\ lines at $1548$ and $1550$~\AA\ that are formed at temperatures around $60$--$250\times10^3$~K and sample the transition region.
There are also considerable contributions from the surrounding continuum because of the broad passbands ($20$ and $275$~\AA, respectively).
The $1700$-\AA\ passband has a width of $200$~\AA, but does not include the \CIV\ lines.
In this analysis, which focuses on the \CIV\ contribution, we employ the technique of
	\citet{1998SoPh..183...29H}
to construct an image sequence in the \CIV\ lines through linear combination of the three bandpasses.
Because TRACE takes images sequentially, we used temporal interpolation of the $1600$ and $1700$-\AA\ images to match the $1550$-\AA\ images.
We prefer to use interpolation over the more usual nearest-neighbor selection, because the averaging inherent to interpolation reduces the noise.
Comparison between the two techniques shows only small differences.

The final step of the reduction is cross-alignment of the sequences of \Ha\ and \CIV\ images.
This was done using several small bright points visible in \Ha\ $-700$~\mA\ as well as the UV continua.
The resulting \CIV\ image sequence was scaled and cut out to match the \Ha\ images.
Since the cadence of the \CIV\ sequence is much faster than that of the \Ha\ sequences, and since the four \Ha\ sequences are staggered in time, we produced four \CIV\ sequences for each of the \Ha\ sequences by nearest-neighbor sampling.
Visual inspection shows good alignment ($<0.5\arcsec$) between the \CIV\ and \Ha\ sequences.

%%%%%%%%%%%%%%%%%%%%%%%%%%%%%%%%%%%%%%%%%%%%%%%%%%%%%%%%%%%%%%%%%%%%%%%%%%%
\section{Analysis and results}\label{sec:analysis}
%%%%%%%%%%%%%%%%%%%%%%%%%%%%%%%%%%%%%%%%%%%%%%%%%%%%%%%%%%%%%%%%%%%%%%%%%%%

% rubber: depend bm_spicsample12.eps
\figuretwo{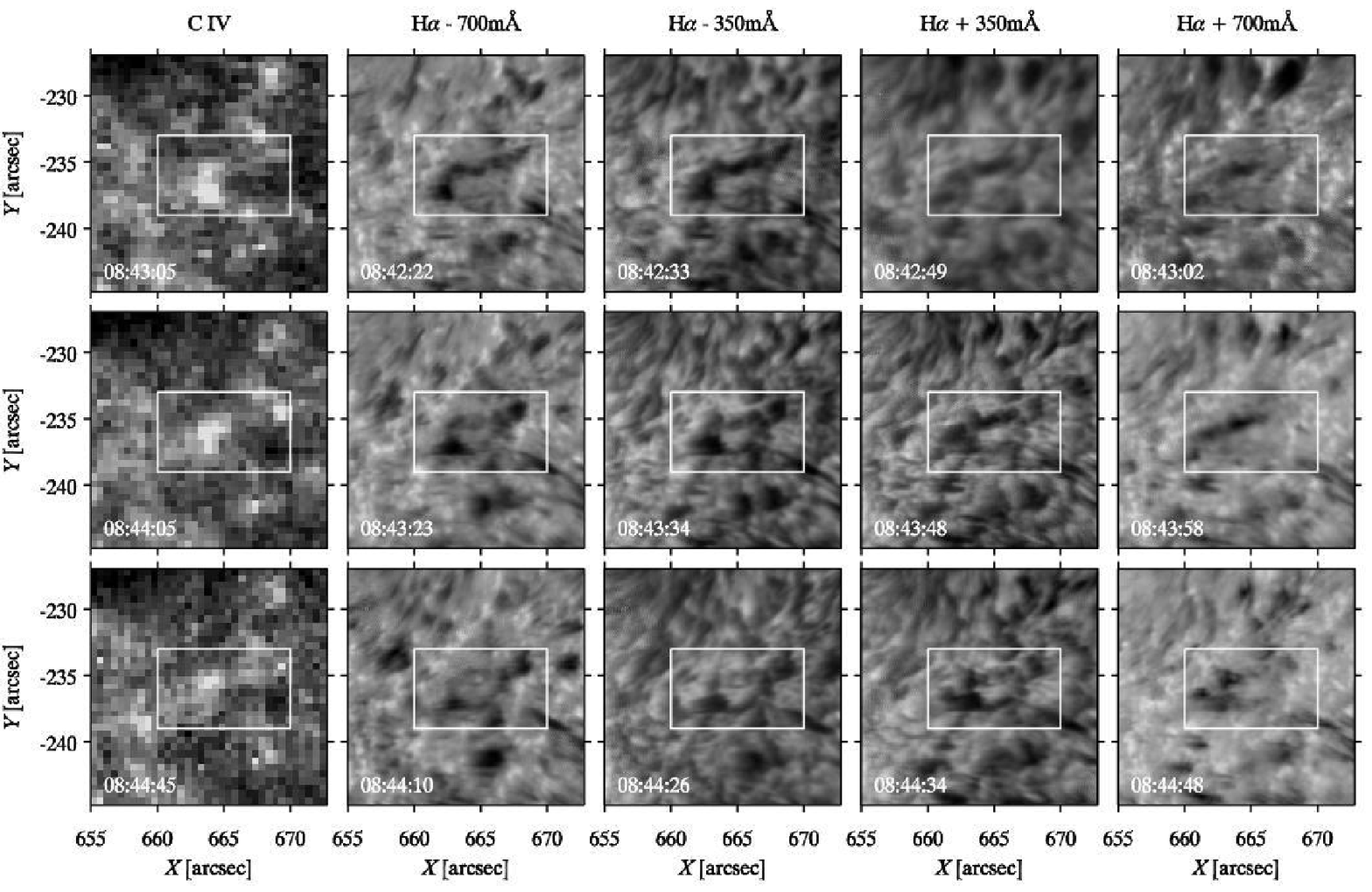}{fig:spicsample12}{%
Temporal evolution of the fibril in the first column of Fig.~\ref{fig:spicsample} for \CIV\ and \Ha\ at the four wavelengths.
For each \Ha\ wavelength, the gray scales of the three panels have been adjusted to make their brightness distributions equal.
The top and bottom rows sample the scene approximately one minute before and after the middle row, as specified by the time of observation per panel.
The white frame outlines the fibril of interest.
It is first most clear in the blue wing of \Ha\ (top row, second and third panels), then moves through the line to the red wing and disappears partially (middle and bottom row, fourth and fifth panels).
It is present in all \Ha\ images, but best visible in the blue wing at first, and later in the red wing.
The \CIV\ panels show diminishing brightness at its location.}

Figure~\ref{fig:sample2003a} shows a cutout of an example $1600$-\AA\ image to illustrate the common field of view between TRACE and the SST.
The solid rectangle outlines the SST field of view, the dashed rectangle outlines the cutout in Fig.~\ref{fig:sample2003b}.
The latter contains active-region plage which appears as mossy plage in a TRACE $171$-\AA\ sequence taken an hour earlier.
It harbors many small, dark \Ha\ features that are best visible in the $\pm700$-\mA\ panels.
They are also present at the inner-core wavelengths but with more confusion rising from feature crowding.
Closer to the limb, such confusion would be much worse due to projection effects and also affect the outer wing panels, but at $\mu=0.66$ single features are well identifiable in these data.
Most of these features display very dynamic behavior in the \Ha\ sequences.
We call these ``dynamic fibrils'' because they come and go on timescales of a few minutes only; they are the subject of this study.
Fibrils that are more stable in appearance and tend to be longer are not the subject of this paper.

% rubber: depend bm_spicsample14.eps
\figuretwo{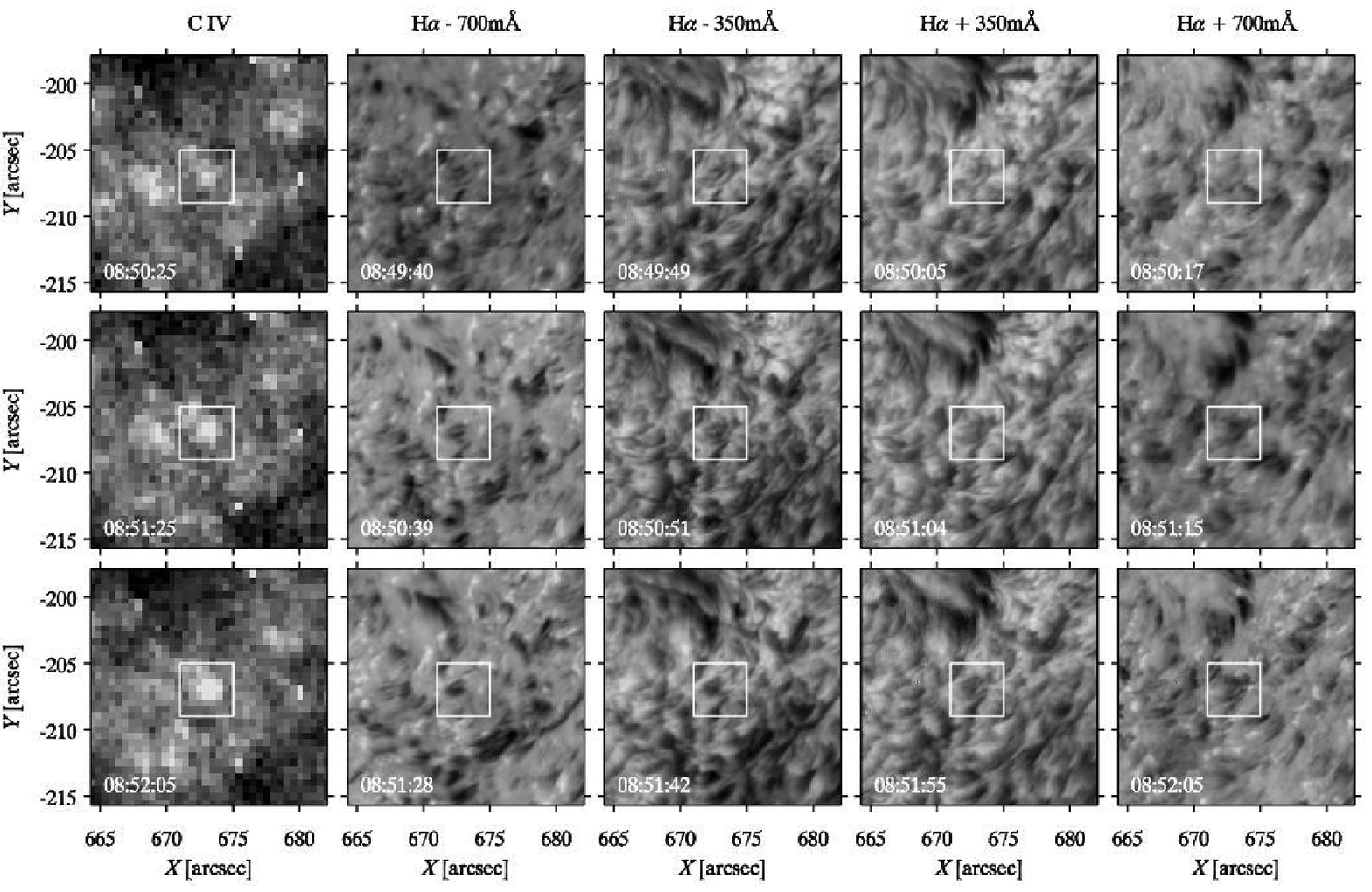}{fig:spicsample14}{%
Temporal evolution of the fibril in the second column of Fig.~\ref{fig:spicsample}, in the same format as Fig.~\ref{fig:spicsample12}.
The central brightening in the \CIV\ panels corresponds to darkening in \Ha\ that progresses from the blue wing (top row, second and third panel) to the red wing (middle and bottom row, fourth and fifth panel).
In the middle row, the dark structure around $(676\arcsec,-212\arcsec)$ expands in the blue wing of \Ha, and migrates toward the red wing while brightening markedly in \CIV\ (bottom row).
Some other structures visible in \Ha\ are not identifiable in \CIV, such as the dark feature around $(678\arcsec,-210\arcsec)$ in the middle \Ha\ $+700$~\mA\ image.}

% rubber: depend bm_spicsample00.eps
\figuretwo{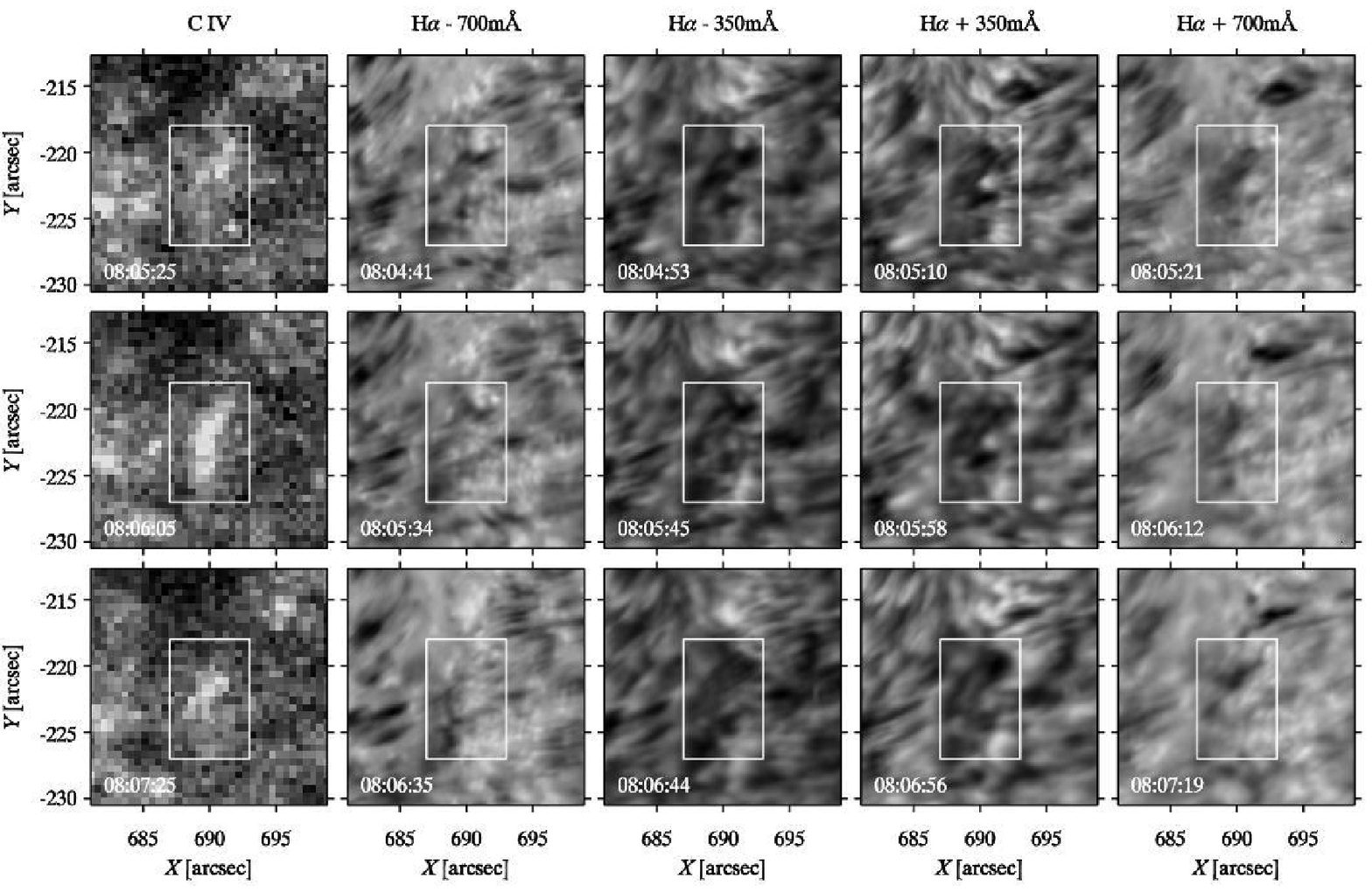}{fig:spicsample00}{%
Temporal evolution of the fibril in the third column of Fig.~\ref{fig:spicsample}, in the same format as Fig.~\ref{fig:spicsample12}.
The bright structure in the \CIV\ panels corresponds to darkening in \Ha\ best visible in the $\pm350$~\mA\ passbands (third and fourth column).
In the top row, only a small feature is present in the far blue wing around $(690\arcsec,-220\arcsec)$ (second panel).
There is already considerable absorption in the far red wing (fifth panel).
The \CIV\ brightening develops quickly and is most prominent in the middle row.
It is morphologically most similar to the absorption in \Ha\ $+700$~\mA\ a minute earlier (top row, fifth panel).
In the bottom row it has largely disappeared.
The fibril around $(690\arcsec,-220\arcsec)$ in the \Ha\ $-700$~\mA\ panel has now progressed into the red wing of \Ha.
Another fibril is present around $(693\arcsec,-215\arcsec)$.
Its associated \CIV\ brightening is less clearly seen and offset by about $1\arcsec$ from the rightmost edge of the absorption in \Ha $+700$~\mA.}

%%%%%%%%%%%%%%%%%%%%%%%%%%%%%%%%%%%%%%%%%%%%%%%%%%%%%%%%%%%%%%%%%%%%%%%%%%%
\subsection{\CIV\ emission from dynamic fibrils}\label{sec:civspic}

We searched for spatio-temporal coincidence of enhanced \CIV\ emission and the occurrence of dynamic \Ha\ fibrils by detailed inspection of the image sequences.
Often, \CIV\ brightness enhancements do not have obvious counterparts in \Ha, but some of them clearly correspond closely to dark dynamical features in the red wing of \Ha, not only by appearing at the same location and close in time, but also by having morphological similarity.
This has earlier been noticed by
	\cite{2003ApJ...590..502D}. % BDP
We identified 16~cases in which the correspondence between a dynamic fibril in \Ha\ and \CIV\ emission is particularly evident.
Their locations are marked by the white diamonds in the leftmost panel of Fig.~\ref{fig:sample2003b}.
Five examples are shown in Fig.~\ref{fig:spicsample}.
The morphological correspondence is obviously imperfect, which in part can be attributed to noise in the UV images (the TRACE CCD has lost considerable sensitivity since its launch), to the relatively poor angular resolution of the \CIV\ construct, to residual seeing in \Ha, and to intrinsic difference in shape.
However, the number of paired features, their morphological similarities, and the fact that they tend to track each other in their temporal evolution give confidence that their identification as pairs is correct.
The intrinsic shape differences are discussed below.

In Figs.~\ref{fig:spicsample12}--\ref{fig:spicsample00}, we present the samples of the first three columns of Fig.~\ref{fig:spicsample} in detail.
Figure~\ref{fig:spicsample12} shows the temporal evolution of the leftmost case in Fig.~\ref{fig:spicsample}.
The fibril appears as a dark structure especially in \Ha\ $-700$ and $-350$~\mA\ (second and third panel in the top row).
It consists of a slender feature extending from $(663\arcsec,-236\arcsec)$ to $(668\arcsec,-234\arcsec)$, and a smaller squarish feature around $(662\arcsec,-237\arcsec)$.
There is already a brightening in the \CIV\ construct (top row, first panel) appearing just to the right of the square feature around $(664\arcsec,-237\arcsec)$.
In the middle row of Fig.~\ref{fig:spicsample12}, the elongated feature has largely disappeared in the blue wing, but has become more clearly visible in the red one.
The square feature is still visible in the blue wing and does not have an appreciable signature in the $+700$-\mA\ passband.
The lower part of the \CIV\ brightening shrinks slightly, but the upper part coincident with the left side of the elongated \Ha\ feature remains bright, while the other end around $(669\arcsec,-234\arcsec)$ brightens.
The bottom row shows the scene about one minute later.
The elongated \Ha\ feature has mostly disappeared, the square feature is now evident throughout \Ha, and the left \CIV\ brightening has shrunk further.

Figure~\ref{fig:spicsample14} similarly details the example in the second column of Fig.~\ref{fig:spicsample}.
In this case, a small fibril appears in \Ha\ $-700$ and $-350$~\mA\ around $(672\arcsec,-207\arcsec)$ (second and third panel, top row).
A brightening in \CIV\ develops quickly at the same location.
The fibril is visible in all \Ha\ passbands in the middle row.
A minute later (bottom row) it is still clearly visible in the far red wing (fifth panel), and appears somewhat more clearly in \Ha\ $-700$~\mA\ (second panel), indicative of recurrent behavior.
It remains clearly visible in the \CIV\ image.
Despite its small size in \Ha, there is considerable brightening in \CIV.

Figure~\ref{fig:spicsample00} shows the example in the third column of Fig.~\ref{fig:spicsample}.
The \Ha\ $\pm350$~\mA\ panels (third and fourth column) suffer much from feature crowding that makes precise identification of features difficult.
However, the similarity between, e.g., the absorption in the \Ha\ $+700$~\mA\ in the top row and the \CIV\ emission in the middle row is striking.
A small feature around $(690\arcsec,-220\arcsec)$ is first clearly visible in the far blue wing (top row, second panel), and later at the same location with similar morphology in the far red wing (bottom row, fifth panel).
It appears at the same location as the absorption in \Ha\ $+700$~\mA\ in the top row (fifth panel), pointing to recurrent behavior.
It also shows associated \CIV\ emission in the middle row.
Another fibril around $(693\arcsec,-215\arcsec)$ develops \CIV\ brightening more slowly and less obviously.
In this case, the \CIV\ emission is offset by about $1\arcsec$ from the rightmost edge of the absorption in \Ha\ $+700$~\mA.

The remaining thirteen cases not detailed here all exhibit similar evolution: a dark absorption feature first appears in the blue wing of \Ha\ and then migrates to the red wing, while corresponding brightening appears in \CIV.
We emphasize, however, that there are many more fibrils in the SST field of view that show such smooth progression through \Ha\ without obvious corresponding \CIV\ emission.

%%%%%%%%%%%%%%%%%%%%%%%%%%%%%%%%%%%%%%%%%%%%%%%%%%%%%%%%%%%%%%%%%%%%%%%%%%%
\subsection{Correlations}\label{sec:pearson}

We computed Pearson correlation coefficients
	\cite[e.g.,][]{numericalrecipes2nd} % Press numerical recipes
for the 16 selected locations between \Ha\ and \CIV\ intensity as a function of time delay.
They describe the amount of linear correlation between the two intensities, i.e., a correlation coefficient $C=0$ indicates a complete lack of linear correlation, while $C=1$ and $-1$ reflect perfect linear correlation and anticorrelation, respectively.
In each case, the fibril area was determined by first selecting those pixels in \CIV\ that are more than two standard deviations above the average value around the fibril, and dilating the result with a $30\times30$-pixel kernel.
These regions are indicated by white outlines in the five examples in Fig.~\ref{fig:spicsample}.
For each \Ha\ wavelength, we next correlated the \CIV\ frame in which the feature is best visible to the five \Ha\ frames straddling it in time.
As to the correlation coefficients $C$ of bright \CIV\ versus dark \Ha\ $+700$~\mA\ derived for the 16~cases, we find that their extremes reach a minimum of $-0.7$ and a maximum of $-0.2$, while their average is $-0.5$.
The delay between \CIV\ and the best match in \Ha\ $+700$~\mA\ varies between $13$ and $-109$~s with an average value of $-33\pm30$~s, where a negative delay indicates that \Ha\ $+700$~\mA\ precedes \CIV.
Despite the large error, the negative delay appears significant because it is found in all but two cases.

% rubber: depend bm_timecorr.eps
\figuretwo{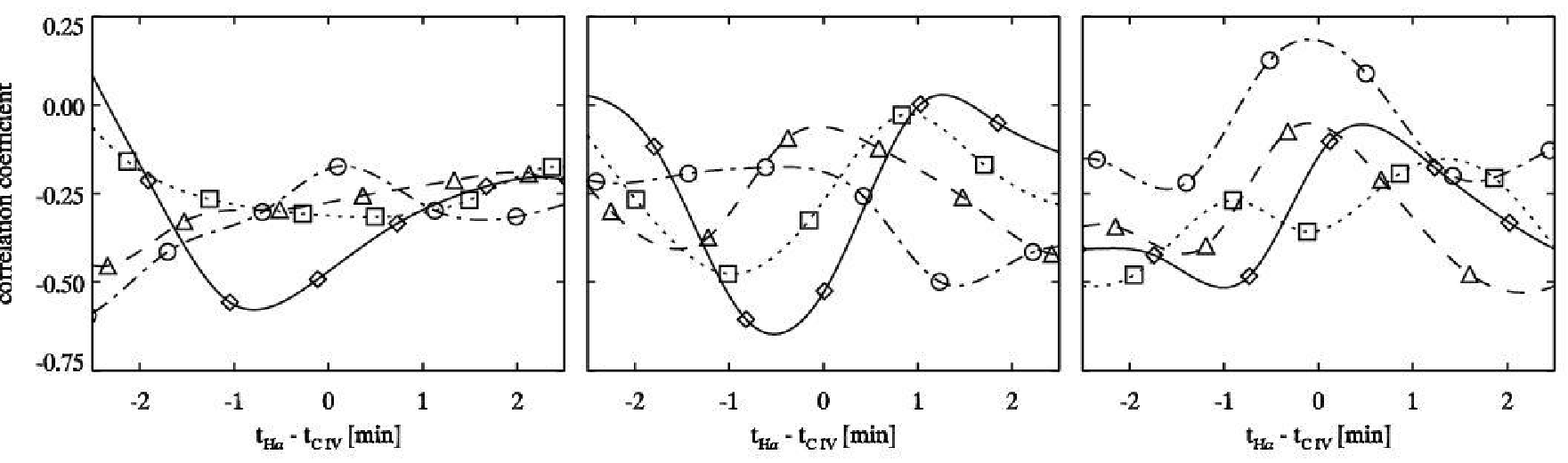}{fig:timecorr}{%
Three samples of the correlation between \Ha\ and \CIV\ intensity as a function of time delay, corresponding to the first three cases in Fig.~\ref{fig:spicsample}, respectively.
Curve coding: correlation between \CIV\ and \Ha\ $-700$~\mA\ (dashed-dotted, circles), \Ha\ $-350$~\mA\ (dashed, triangles), \Ha\ $+350$~\mA\ (dotted, squares), and \Ha\ $+700$~\mA\ (solid, diamonds).}

Figure~\ref{fig:timecorr} shows the correlations against time for the first three examples of Fig.~\ref{fig:spicsample}.
For all fibrils the anticorrelation is most obvious for the $+700$-\mA\ passband, and varies for the other passbands.
The fibrils show a similar time delay of about one minute for the $+700$-\mA\ passband, in the sense that \CIV\ brightens later than the red wing darkens.
Especially the fibril in the center panel shows temporal progression from blue to red across the line.
It also shows recurrent behavior: the correlation drops again at large time difference, at first in the blue wing, and later in the red.
The fibril in the rightmost panel displays similar behavior.
Figure~\ref{fig:histograms} displays histograms of the delays of best anticorrelation between \Ha\ and \CIV\ intensity for each \Ha\ wavelength.
It shows a clear peak at negative values for the best anticorrelations between \Ha\ $+700$~\mA\ and \CIV\ brightness, indicating that the half-minute negative delay is significant despite a large error, and less so between $+350$~\mA\ and \CIV\ brightness.
The histograms of delays for the best anticorrelation between \Ha\ blue wing and \CIV\ brightness appear featureless.
Large positive delays can be interpreted as signs of recurrent behavior.

%%%%%%%%%%%%%%%%%%%%%%%%%%%%%%%%%%%%%%%%%%%%%%%%%%%%%%%%%%%%%%%%%%%%%%%%%%%
\section{Discussion}\label{sec:discussion}
%%%%%%%%%%%%%%%%%%%%%%%%%%%%%%%%%%%%%%%%%%%%%%%%%%%%%%%%%%%%%%%%%%%%%%%%%%%

The observations presented here add temporal evolution to the study by
	\cite{2003ApJ...590..502D}.
In a one-hour multi-wavelength sequence, we have identified 16 dynamic \Ha\ fibrils with associated \CIV\ emission.
In all cases, the corresponding dark \Ha\ feature progresses in a smooth, continuous fashion from the blue wing to the red one in a few minutes, and only then shows the best correlation with the \CIV\ brightness feature.
It often also displays recurrent behavior.
Many \Ha\ fibrils appear to have some corresponding \CIV\ emission, though it is conspicuous in only a few cases.

Another aspect of interest is the width of the \Ha\ line in dynamic fibrils.
Inspection of the \Ha\ sequence shows that most, if not all, short-lived \Ha\ fibrils show such progression from the blue to the red wing, but most are not obviously associated with \CIV\ brightening.
A case in point is the dark feature around $(678\arcsec,-210\arcsec)$ in Fig.~\ref{fig:spicsample14}.
It is present in the $\pm350$~\mA\ panels in the top row, and shows progression to the red wing (second row, $+350$ and $+700$~\mA\ panels) without clear \CIV\ brightening.

We interpret the progression as to result from matter first moving up in the direction of the observer, and then back toward the surface.
This interpretation agrees well with the description of active-region moss as jets moving upward in the atmosphere and periodically obscuring the hot footpoints of loops
	\citep{1999SoPh..190..409B, % Berger moss
	1999SoPh..190..419D, % De Pontieu moss at high t res
	2000ApJ...537..471M}. % Martens moss = hot footpoints + jets
We do not have concurrent observations in TRACE $171$~\AA\ or $195$~\AA, so we cannot establish that our dynamic \Ha\ fibrils correspond to such obscuration in mossy plage.
However, since the relationship between moss and chromospheric jets is well-established
	\cite[e.g.,][]{2003ApJ...590..502D}, % De Pontieu, moss vs C IV vs Halpha
it is likely that the dynamic fibrils described here correspond to the jets that cause moss.

% rubber: depend bm_histograms.eps
\figureone{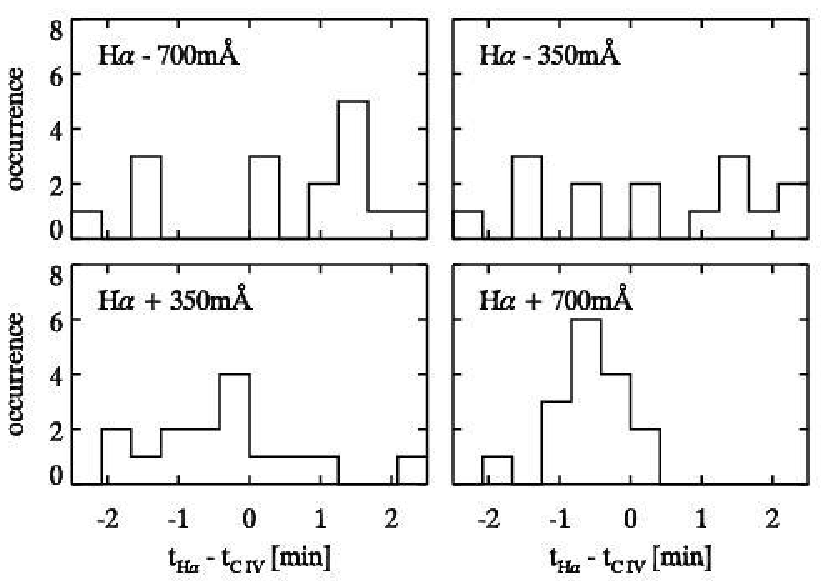}{fig:histograms}{%
Histograms of the time delay of the best anticorrelation between \Ha\ and \CIV\ intensity for each \Ha\ wavelength.}

The observation that the \CIV\ emission occurs only after the fibril has progressed to the red wing of \Ha, i.e., when the matter returns toward the solar surface, implies that the \CIV\ emission line itself should be shifted to the red as well.
Spectroscopic observations indeed show that this is the case
	\citep{1986ApJ...305..947D}. % Dere et al on disk EUV jets

The curves in Fig.~\ref{fig:timecorr} show that darkness in \Ha\ $+700$~\mA\ correlates well with \CIV\ brightness occurring half a minute later.
The disappearance of the \Ha\ absorption at that time may be due to the narrow width of the SOUP passband ($130$~\mA), which registers only a narrow range of velocities.
If the fibril still accelerates downward while being heated, it may reach the temperature needed for \CIV\ emission while shifting out of the SOUP passband.

As already noted, not all dynamic \Ha\ fibrils show corresponding \CIV\ emission.
Some may be missed when they evolve faster than our temporal sampling, which is limited by the $\sim$$1$~minute cadence of the \Ha\ sequences.
Some others may be missed due to lack of instrument sensitivity.

What precisely is the origin of this \CIV\ emission that apparently corresponds to cool \Ha\ structures?
The emission may originate from a hot sheath around a cool fibril acting as interface to the surrounding corona, as proposed by 
	\cite{1983ApJ...267..825W}. % Withbroe EUV limb emission =? spicules
Such a sheath could be formed through cross-field thermal conduction
	\citep{1990ApJ...362..364A}, % Athay 
and would be expected to produce \CIV\ emission that is cospatial with the \Ha\ absorption.
Estimated radiative losses show that only a very thin layer of hot material is required to explain the observed intensity in our TRACE images, so that cross-field conduction cannot be ruled out on grounds of effectiveness or timescale.
However, we get the impression from our observations that many of the brightness enhancements in \CIV\ are systematically shifted toward the upper end point of the corresponding \Ha\ fibril.
This is the case for all 16~fibrils studied here; in some the \CIV\ brightening extends as far out as $1\arcsec$ beyond the apex of the feature in \Ha\ $+700$~\mA.
Co-spatial correlations have been computed in Sect.~\ref{sec:pearson}.
The temporal delay shown by the curves in Fig.~\ref{fig:timecorr} is not invalidated by the observation that the \Ha\ and \CIV\ features are not co-spatial.
It even serves as confirmation: the offset \CIV\ emission is more strongly correlated with the \Ha\ fibril in its appearance of, say, half a minute ago, when it had a larger extent.
	\cite{1986ApJ...305..947D} % Dere et al on disk EUV jets
also noted that in quiet-sun observations the most intense \CIV\ emission is produced just beyond \Ha\ mottles.
This finding suggests that there is a large temperature gradient along the structure, with a cold footpoint in the chromosphere and a hot top in the corona.
The energy required to heat the tip of the fibril may then be provided by the corona, possibly through parallel heat conduction, or by hot coronal material filling the wake of the descending fibril.
	\cite{1987SoPh..114..223D}
derived extremely small filling factors from HRTS spectra, and concluded that UV fibrils are sparsely filled with numerous hot microscopic structures.
Our observations cannot confirm or disprove this, other than that these microscopic structures should develop toward the end of the fibril's life, during or after the descent of the fibril back to the lower atmosphere, in order to reproduce the observed time delay between the appearance in the far red wing and the \CIV\ emission.
It is unclear what mechanism may cause heating in such a highly inhomogeneous spatial pattern, or why it would preferentially cause heating during the downward contraction of the fibril.

Another possibility is that a density enhancement occurs when the jet descends, causing brightness enhancements in optically thin conditions, as is the case for the \CIV\ lines used here.
Lastly, coronal and transition-region EUV radiation may ionize a sheath, causing \CIV\ emission through recombination.

Classically, spicular line widths and profiles of \Ha\ and \CaIIHK\ pose inconsistencies between thermal and non-thermal broadening
	\citep[e.g.,][]{1968SoPh....3..367B}. % Beckers, spicule reviews
Our 16~specimens show that \Ha\ darkening often occurs simultaneously in all four \Ha\ passbands, with the most pronounced darkening progressing from the blue to the red wing.
This is also shown by the correlation curves in Fig.~\ref{fig:timecorr}.
It requires large line broadening which may be caused by unresolved structure, non-thermal motions including turbulence, or a high temperature.
The latter would be consistent with associated \CIV\ emission, though contradicting the common belief of $\sim$$10^4$~K temperature with large non-thermal broadening.
It is of interest to map chromospheric line widths at high resolution in this context.

%%%%%%%%%%%%%%%%%%%%%%%%%%%%%%%%%%%%%%%%%%%%%%%%%%%%%%%%%%%%%%%%%%%%%%%%%%%
\section{Conclusions}\label{sec:conclusion}
%%%%%%%%%%%%%%%%%%%%%%%%%%%%%%%%%%%%%%%%%%%%%%%%%%%%%%%%%%%%%%%%%%%%%%%%%%%

We have identified 16 dynamic \Ha\ fibrils in high-resolution image sequences which show corresponding \CIV\ emission in the UV.
Without exception, these fibrils show a progression of the absorption feature through \Ha\ from the blue to the red wing, often with recurrent behavior.
Many, if not all, dynamic fibrils that do not display identifiable \CIV\ emission also show this progression.
We interpret it as matter flowing first up and then down in the solar atmosphere.

We conclude that some plasma associated with dynamic fibrils is heated to transition-region temperatures.
The \CIV\ emission appears at the apex of the \Ha\ fibril, sometimes extending beyond it.
It best resembles the far red \Ha\ wing of about half a minute earlier.
We attribute this apparent delay to the offset of the \CIV\ emission beyond the tip of the \Ha\ fibril.

Whichever heating mechanism operates, our observed correlation between \Ha\ and \CIV\ fibrils clearly provides constraints on theoretical models.
Detailed comparisons with advanced 2D or 3D numerical simulations of the low solar atmosphere will be necessary to determine which effect dominates the heating of fibril plasma to $10^5$~K.

%%%%%%%%%%%%%%%%%%%%%%%%%%%%%%%%%%%%%%%%%%%%%%%%%%%%%%%%%%%%%%%%%%%%%%%%%%%
\begin{acknowledgements}
We are indebted to R.J.~Rutten for many fruitful discussions.
We thank the anonymous referee for suggesting many improvements.
AdW thanks P.~Martens and H.~Winter for insightful comments.
AdW acknowledges travel support from the Leids Kerkhoven-Bosscha Fonds and hospitality at LMSAL and MSU.
BDP was supported by NASA grants NAG5-11917, NNG04-GC08G and NAS5-38099 (TRACE).
\end{acknowledgements}

%%% BIBLIOGRAPHY %%%%%%%%%%%%%%%%%%%%%%%%%%%%%%%%%%%%%%%%%%%%%%%%%%%%%%%%%%

\end{document}